\documentclass[runningheads]{llncs}
\usepackage{graphicx}
\usepackage{amsmath}
\usepackage{hyperref}
\usepackage{array}
\usepackage{tabularx}
\usepackage{amssymb}
\usepackage{tablefootnote}
\usepackage{lmodern}
\usepackage[T1]{fontenc}
\usepackage{bm}
\usepackage{bbm}
\usepackage{booktabs}
\usepackage[linesnumbered,ruled,vlined]{algorithm2e}
\usepackage{cite}
\SetKwInput{kwInputs}{Inputs}

\begin{document}

%
\title{Finding Near-Optimal Portfolios With Quality-Diversity}
%
%
\author{Bruno Ga\v{s}perov \and
Marko \DJ urasevi\'c \and
Domagoj Jakobovic}
\authorrunning{B. Ga\v{s}perov et al.}
%
\institute{Faculty of Electrical Engineering and Computing, \\ University of Zagreb, Zagreb, Croatia
\\
\email{\{bruno.gasperov,marko.durasevic,domagoj.jakobovic\}@fer.hr}}

\maketitle              
\begin{abstract}
The majority of standard approaches to financial portfolio optimization (PO) are based on the mean-variance (MV) framework. Given a risk aversion coefficient, the MV procedure yields a single portfolio that represents the optimal trade-off between risk and return. However, the resulting optimal portfolio is known to be highly sensitive to the input parameters, i.e., the estimates of the return covariance matrix and the mean return vector. It has been shown that a more robust and flexible alternative lies in determining the entire region of near-optimal portfolios. In this paper, we present a novel approach for finding a diverse set of such portfolios based on quality-diversity (QD) optimization. More specifically, we employ the CVT-MAP-Elites algorithm, which is scalable to high-dimensional settings with potentially hundreds of behavioral descriptors and/or assets. The results highlight the promising features of QD as a novel tool in PO.

\end{abstract}
\section{Introduction}
\subsection{Portfolio optimization and near-optimal portfolios}

Portfolio optimization (PO) entails finding the optimal allocation of limited capital across a range of available financial assets within a specified timeframe. The optimality is typically defined with respect to a risk-adjusted return metric, such as the Sharpe ratio \cite{sharpe}, or other metrics that account for the investor's specific risk preferences, like the CARA utility function \cite{babcock}. Classical approaches to PO are deeply rooted in the mean-variance (MV) framework, also referred to as modern portfolio theory (MPT) \cite{markowitz}. MV optimization provides a systematic methodology for constructing optimal portfolios that leverage the principle of diversification. By distributing investments among assets with dissimilar risk-return profiles (e.g., assets with mutually uncorrelated or negatively correlated returns), individual risks offset each other, thereby reducing the total portfolio risk. Given a risk aversion coefficient, the MV optimization process diversifies assets to produce a single optimal portfolio representing the best risk-return trade-off. The set of optimal portfolios for different risk preferences constitutes a Pareto front called the efficient frontier.  

However, the MV framework is predicated upon a number of simplified and unrealistic assumptions, including the normality of returns and the stationarity of the return covariance matrix. It also tends to produce portfolios highly concentrated in only a few assets, jeopardizing diversification. Furthermore, MV optimizers are particularly sensitive to estimation errors, with small changes in input parameters, especially expected return estimates, noticeably affecting the resulting optimal portfolio weights \cite{best}. Such notorious issues have spurred researchers to propose novel PO approaches focusing on improving robustness. Various attempts have been made in this direction, including methods based on shrinking the covariance matrix \cite{ledoit} or expected returns \cite{black}, introducing constraints to the original MV framework \cite{demiguel}, resampling the efficient frontier \cite{michaud}, or directly applying techniques from the area of robust optimization \cite{yin}. 

Another alternative, the focus of our work, involves identifying portfolios that are not strictly optimal but rather near-optimal (with respect to MV optimality). In the first phase (the optimization process), an entire subspace of mutually diverse\footnote{In terms of their distance in the space of admissible portfolio weights or, more generally, some behavior space.} near-optimal portfolios is determined without focusing on any specific portfolio. This offers enhanced robustness to estimation errors \cite{degraaf}, as it sidesteps the complex issues stemming from the interplay between the objective and the inputs, which are inherent to the MV optimization. In the second phase (the \textit{a posteriori} analysis), the investor selects the final portfolio from the subspace of near-optimal portfolios. This decision-making scheme allows investors to incorporate their expert opinions, subjective views, or any other soft factors that may be challenging to integrate directly into an optimization problem formulation. Simultaneously, it permits the \emph{ad-hoc} consideration of market frictions (e.g., transaction costs) and other regulatory, liquidity, and risk concerns, which are crucial in real-life PO but were not necessarily applied as constraints during the first phase. For instance, when rebalancing portfolio weights, investors may prefer a near-optimal portfolio similar to the current one over the MV optimal portfolio due to the former's lower turnover, and hence also lower transaction costs. Therefore the consideration of near-optimal portfolios presents a fruitful opportunity for improving the process of PO.

\subsection{Prior research}

\subsubsection{Near-optimal portfolios}
The problem of identifying and analyzing near-optimal portfolios has been scrutinized in several studies. Van der Schans and de Graaf \cite{schans} pioneered a novel methodology for constructing such portfolios, outlined as follows. Let $\mathcal{F}: \mathcal{W} \rightarrow \mathcal{Z}$ be the mapping from the set of admissible portfolio weights $\mathcal{W}$ to the risk-return space $\mathcal{Z}$. First, an optimal portfolio $\boldsymbol{w_0}$ (satisfying $\mathcal{F}(\boldsymbol{w_0}) \in \mathcal{E}$ where $\mathcal{E} \subset \mathcal{Z}$ is the efficient frontier) is selected as a reference portfolio in accordance with the investor's risk preferences. A small region $\mathcal{R} \subset \mathcal{Z}$ 
around $\mathcal{F}(\boldsymbol{w_0})$ is then defined and the portfolio $\boldsymbol{w_1}$, s.t. $\mathcal{F}(\boldsymbol{w_1}) \in \mathcal{R}$, located furthest away\footnote{In the sense of the Euclidean distance between the portfolio weight vectors $||\boldsymbol{w_1}-\boldsymbol{w_0}||$.} from $\boldsymbol{w_0}$, is found. Subsequently, the portfolio $\boldsymbol{w_2}$, again s.t. $\mathcal{F}(\boldsymbol{w_2}) \in \mathcal{R}$, positioned furthest away from the convex hull $\mathcal{H}$ spanned by previously found portfolios $\{\boldsymbol{w_0}$, $\boldsymbol{w_1}\}$ is itself added to $\mathcal{H}$. Generally, in step $i$:

\begin{equation}
\boldsymbol{w_i}=\underset{\mathcal{F}(\boldsymbol{w}) \in \mathcal{R}}{\arg \max } \ d\left(\boldsymbol{w}, \mathcal{H} \right),
\end{equation}
with $d\left(\boldsymbol{w}, \mathcal{H} \right) =\min _{\boldsymbol{w'} \in \mathcal{H}}\left\|\boldsymbol{w}-\boldsymbol{w'}\right\|$. This process iterates until $\mathcal{S} \subset \mathcal{Z}$, the convex hull in the risk-return space corresponding to the portfolios in $\mathcal{H}$, covers $\mathcal{R}$ to the required level of precision $\epsilon$. Finally, a set of $K$ diverse near-optimal portfolios $\{\boldsymbol{w_0}, \boldsymbol{w_1}, \ldots, \boldsymbol{w_{K-1}} \}$ is obtained. Since a convex combination of near-optimal portfolios is also near-optimal \cite{degraaf}, any portfolio in $\mathcal{H}$ acts as a viable option for the investor. The authors finally show that this region of near-optimal portfolios is more robust to input estimate uncertainties than a single optimal portfolio. However, this approach exhibits several limitations. Firstly, it involves solving a difficult non-convex optimization problem of finding the portfolio furthest from the convex hull, leading the authors to use a somewhat \emph{ad hoc} combination of a support vector machine \cite{wang} and a basin-hopping algorithm \cite{wales}. Secondly, it becomes unfeasible in high-dimensionality settings, i.e., when faced with a large number of assets ($N$), which is typically the case in modern PO\footnote{Van Eeghen \cite{eeghen} reports computation times of around $2$ hours and more per run already for $N>20$.}. Thirdly, it is restricted to finding near-optimal portfolios directly in the space of portfolio weights and is not easily applicable to other possible behavior (feature) spaces, which may comprise variables such as fundamentals, technicals, and risk factors. Put differently, it does not tackle the generalized variant where in step $i$:
\begin{equation}
\boldsymbol{w_i}=\underset{\mathcal{F}(\boldsymbol{w}) \in \mathcal{R}}{\arg \max } \ d\left(\phi(\boldsymbol{w}), \mathcal{H} \right)
\end{equation}
where $\mathcal{H} = \{\phi(\boldsymbol{w_0})$, \ldots, $\phi (\boldsymbol{w_{i-1}})\}$ for some function $\phi$. To ameliorate some of these problems, Cajas \cite{cajas} proposed the near-optimal centering (NOC) method, based on finding the analytic centers of near-optimal regions. The method can be used in conjunction with any convex risk measure and has been empirically demonstrated to lead to improved diversification and robustness when compared to traditional PO methods. Van Eeghen \cite{eeghen} expands on \cite{schans} by using polytope theory to inspect the structure and robustness of near-optimal regions. Moreover, the author proposes a new implementation of the method from \cite{schans} that reduces computation time while maintaining accuracy.

The topic of near-optimal portfolios has been investigated or touched upon in several other works as well. Chopra \cite{chopra} uses a grid search to discover a subset of near-optimal portfolios and analyzes their composition. Benita \emph{et al.} \cite{benita} emphasize that near-optimal portfolios might provide a higher degree of robustness to various scenarios due to significant differences in their makeups and provide an illustrative example. Lastly, Fagerström and Oddshammar demonstrate that the Conditional Value-at-Risk (CVaR) optimization model tends to produce portfolios that are near-optimal under the MV framework \cite{fagerstrom}, i.e., located very close to the efficient frontier. 

\vspace{-0.5cm}

\subsubsection{Evolutionary computation and quality-diversity optimization} Evolutionary computation (EC) methods have a rich history of successful applications in PO \cite{brabazon}, partly owing to their ability to handle non-convex search spaces that arise when real-world constraints (e.g., buy-in thresholds, turnover constraints, other regulatory and risk constraints) are imposed into the problem setting \cite{branke,qi}. However, there is a significant lack of research regarding the applications of exploration algorithms, such as quality-diversity (QD) \cite{chatz} and novelty search (NS) \cite{lehman} in PO and quantitative finance generally, despite their significant potential and their successes in other domains \cite{gomes2013evolution, pugh2016quality}. The links between portfolio diversification and the divergent search paradigm, while arguably natural, remain understudied. Several somewhat related works nevertheless exist. Zhang \emph{et al.} \cite{zhang} address the specific problem of finding formulaic alpha factors that can predict and explain asset returns, making them suitable for use with multi-factor asset pricing models. To efficiently explore the space of formulaic alphas, with a focus on less frequently visited regions, they propose a search that combines QD and principal component analysis (PCA). This PCA-enhanced search is then used as an integral part of their \verb|AutoAlpha| hierarchical evolutionary algorithm. Another approach is put forth by Yuksel \cite{yuksel}, in which meta-learning QD optimization is employed to tackle the problem of large-scale sparse index tracking. It is concluded that the proposed method can be utilized in other scenarios where diversity among co-optimized solutions is needed, as well as in the presence of noisy reinforcement learning rewards.

\vspace{-0.3cm}

\subsection{Objectives and contributions}

In this paper, we tackle the problem of identifying a diverse set of near-optimal portfolios (i.e., portfolios with risk-return profiles located close to that of the reference optimal portfolio), as part of the broader PO problem. Specifically, we aim to answer the following research question (\textbf{Q}): \textit{How to obtain a wide range of portfolios that are all near-optimal but mutually diverse in the portfolio weight space or the otherwise defined behavior space (BS)?} While previous research has found that, in some tasks, elite solutions are concentrated within a small part of the genotypic space ("the elite hypervolume") \cite{vassila}, our task runs in the opposite direction, as it involves finding genotypically different solutions (portfolios) that are all elite. We set out to test the following hypothesis (\textbf{H}): \textit{The combination of convergent and divergent search provided through QD algorithms can be leveraged to obtain a set of diverse near-optimal portfolios.} The hypothesis stems from the observation that QD algorithms provide a natural choice for the underlying problem due to their ability to yield a range of diverse yet high-performing solutions. While some related approaches exist \cite{zhang,yuksel}, to the best of our knowledge, this paper is the first to approach MV-based PO via QD optimization. To ensure the scalability to high-dimensional behavioral and/or asset spaces, which are ubiquitous in modern finance, the approach is powered by the \verb|CVT-MAP-Elites| algorithm, as vanilla \verb|MAP-Elites| faces the curse of dimensionality. We first use a toy example with only three assets to show that the approach is competitive against a similar approach based on the construction of convex hulls \cite{schans}, and later extend our investigation to a higher-dimensional setting. As will be shown, the experimental results collectively clearly point to the promising capabilities of QD as a novel tool in the arsenal of modern PO practitioners.

\section{Methodology}

\subsection{Problem formulation}
Let $\mathcal{W}$ be the set of all admissible portfolio weights:

\begin{equation}
    \mathcal{W} = \left\{ (w_1, \ldots, w_i, \ldots, w_N) \mid w_i \geq 0 \quad \forall i, \quad \sum_{i=1}^{N} w_i = 1 \right\},
\label{eq3}
\end{equation}
where $N \geq 2$ is the total number of assets and $w_i$ denotes the portfolio weight in the $i$-th asset. For simplicity, short selling is not permitted, but it can be easily integrated into the framework if needed, by relaxing the $w_i \geq 0$ constraint.
Also, let  $b : \mathcal{W} \rightarrow \mathcal{B}$ be a behavior function, mapping $\mathcal{W}$ to a BS denoted by $\mathcal{B}$. Assume that $\mathcal{B}$ is split into $M$ niches, i.e., $\mathcal{B} = \mathcal{N}_1 \cup \mkern3mu \mathcal{N}_2 \cup \mkern3mu \ldots \cup \mkern3mu \mathcal{N}_M$. Finally, let $f: \mathcal{W} \rightarrow \mathbb{R}$ be a fitness function. Each candidate portfolio $\boldsymbol{w}$ is then associated with its behavioral descriptor (BD) $\boldsymbol{b}_{\boldsymbol{w}} \in \mathcal{B}$ and fitness value $f(\boldsymbol{w}) \in \mathbb{R}$. The goal is to find: 

\begin{equation}
\forall \mathcal{N}_i \quad \ \boldsymbol{w}^*=\underset{\boldsymbol{w},\  \boldsymbol{b}_{\boldsymbol{w}} \in \mathcal{N}_i}{\arg \max } \ f(\boldsymbol{w}).
\end{equation}

\subsection{Behavior function and space}

We explore two different behavior functions and BS designs. In both cases, centroidal Voronoi tessellation (CVT) is used to partition the BS into niches. 
\subsubsection{Portfolio weights}In the simplest variant, the behavior function \(b_1\) is set as an identity function, i.e., \(\forall \boldsymbol{w}, \ b_1(\boldsymbol{w})=\boldsymbol{w},\) with \(\mathcal{B}_1 = \mathcal{W}.\) Portfolio weight vectors $\textbf{w}$ simultaneously serve as both genotypes and phenotypes. Similar can be seen in some of the approaches used to tackle the Rastrigin function benchmark \cite{digalakis} through QD algorithms \cite{bossens,sfikas}. 

\subsubsection{Asset's fundamentals} In another variant, we separate the two spaces (genotypic and phenotypic) and set $b_2(\boldsymbol{w})=\boldsymbol{p}_{\boldsymbol{w}}$, where $\boldsymbol{p}_{\boldsymbol{w}} \in \mathcal{B}_2$ is a vector that describes the fundamental properties of the assets that dominate $\boldsymbol{w}$ weight-wise. More specifically:
\begin{equation}
    \mathcal{B}_2 = \left\{ (s_1, \ldots, s_i, \ldots, s_L, c) \mid s_i \geq 0 \quad \forall i, \quad c > 0, \quad \sum_{i=1}^{L} s_i = 1 \right\},
\label{eq1.5}
\end{equation}
where $s_i$ denotes the sector exposure of a portfolio to sector $i$, and $L$ is the number of sectors. Sector exposure is defined as the sum of portfolio weights assigned to assets belonging to the respective sector, i.e., $s_i = \sum_{\substack{j \in S_i}} w_j$, with $S_i$ denoting the set of indices of assets belonging to sector $i$. The variable $c$ denotes normalized market capitalization. Note that, unlike in the previous case, it is not possible to uniformly sample from the BS directly. Also, any other asset characteristics or factors of importance to the investor (e.g., ESG\footnote{Environment, social and governance.} factors, geographical diversification, etc.) might be used instead.

\subsection{Fitness functions}

\verb|Fitness1| - The fitness (quality) function can be given by the negative distance between the risk profile of the candidate portfolio $\boldsymbol{w}$ and that of the reference optimal portfolio $\boldsymbol{w_0}$ (obtained via MV optimization):

\begin{equation}
    f_1(\boldsymbol{w}) = - ||\mathcal{F}(\boldsymbol{w_0})-\mathcal{F}(\boldsymbol{w})|| = -||(\mu_0, \sigma_0)-(\mu, \sigma)||,
\end{equation}
where $(\mu, \sigma) \in \mathcal{Z}$ is a vector consisting of the expected return and the volatility of the portfolio $\boldsymbol{w}$ (and equivalently for $\boldsymbol{w_0}$). 
\\
\noindent
\verb|Fitness2| - Alternatively, a modification of \verb|Fitness1| is proposed:

\begin{equation}
    f_2(\boldsymbol{w})= 
\begin{cases}
    -||(\mu_0, \sigma_0)-(\mu, \sigma)||,& \text{if } (\mu, \sigma) \ \text{not in}\ \mathcal{R} \ \\
    ||\boldsymbol{w}-\boldsymbol{w_0}||,              & \text{otherwise}
\end{cases}
\end{equation}
where $\mathcal{R}$ is the region of near-optimality around $(\mu_0, \sigma_0)$, $\mathcal{R} \subset \mathcal{Z}$, defined as:
\begin{equation}
    \mathcal{R} = \left\{(\mu, \sigma) \mid \mu \geq (1-c) \mu_0, \ \sigma \leq (1+c) \sigma_0 \right\}
\label{eq2}
\end{equation}
for some constant $c \in (0, 1) $. \verb|Fitness2| promotes solutions in each niche whose risk profiles are as close as possible to the risk profile of $\boldsymbol{w_0}$ until close proximity is reached. Once inside the near-optimality region $\mathcal{R}$, solutions that are furthest weight-wise from $\boldsymbol{w_0}$ are preferred to ensure more genotypic diversity through an additional mechanism. Ideally, the distance from the convex hull of the entire archive of near-optimal solutions (instead of only from $\boldsymbol{w_0}$) should be used as the measure. However, for computational efficiency, we employ this simple heuristic which will be shown to demonstrate strong performance. 

\subsection{Recombination operator}

Given the conditions in Eq. \ref{eq3}, a suitable constraint-preserving reproduction operator is required. To this end, a recombination operator (with mutation) that also includes clipping and normalization is used. It is described in great detail in Alg. \ref{algo:newportfolio}. Additional constraints, such as cardinality restrictions or buy-in thresholds, can be incorporated as needed.

\begin{algorithm}[t!]
\SetAlgoLined
\kwInputs{$\text{Parent portfolios} \ \boldsymbol{w_1}, \boldsymbol{w_2}, \text{mutation rate}\ m $}
\KwResult{$\text{Child portfolio} \  \boldsymbol{w_3}$}
$\lambda \gets$ \verb|UniformRandom|[0, 1] \ \tcp*{Randomly generated weight parameter}
$\boldsymbol{\delta} \gets$ \verb|UniformRandom|[$-m, m, \verb|len|(\boldsymbol{w_1})$] \ \tcp*{Mutation vector }
$\boldsymbol{w_3} \gets \lambda \boldsymbol{w_{1}} + (1 - \lambda) \boldsymbol{w_{2}} + \boldsymbol{\delta}$ \tcp*{Child portfolio}
$\boldsymbol{w_3} \gets$ \verb|Clip|($\boldsymbol{w_3}$, 0, 1) \ \tcp*{Clipping to ensure non-negative weights}
$\boldsymbol{w_3} \gets \dfrac{\boldsymbol{w_3}}{\sum_{i=1}^{\text{len}(\boldsymbol{w_3})} \boldsymbol{w_{3,i}}}$ \ \tcp*{Normalization of the child portfolio}
\Return{$\boldsymbol{w_3}$}
\caption{Recombination operator}
\label{algo:newportfolio}
\end{algorithm}

\subsection{Algorithm}

Due to the continual increase in the number of investable securities in financial markets, modern PO typically operates in high-dimensional settings, potentially encompassing hundreds or even thousands of financial assets. The ensuing high-dimensionality can be tackled with the \verb|CVT-MAP-Elites| algorithm, where the number of niches is constant and independent of the dimensionality of the BS \cite{vassiliades}. 
Unless noted otherwise, unstructured archive $\mathcal{A}$ is assumed. For smaller BS dimensionalities (typically 2D to 6D \cite{vassiliades}), the basic \verb|MAP-Elites| algorithm \cite{mouret} can be used instead. We use \verb|CVT-MAP-Elites| in all of the performed experiments. 

\section{Experimental results}

\subsection{Toy example}

\begin{figure}[ht!]
\includegraphics[width=8cm]{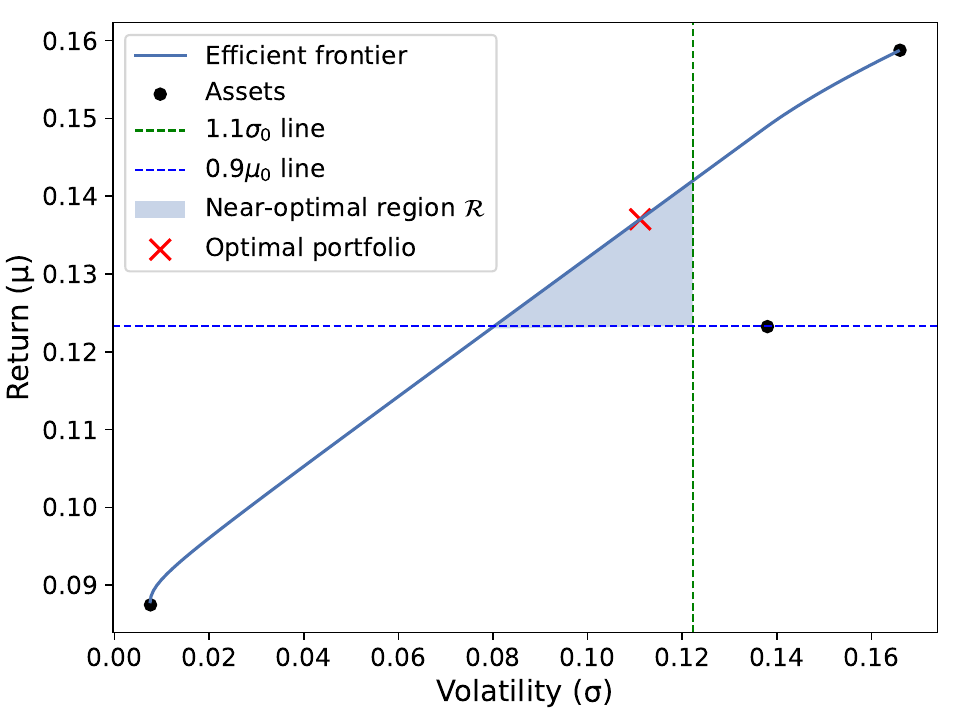}
\centering
\caption{The MV efficient frontier with the chosen optimal portfolio and its region of near-optimality.}
\label{figure1}
\end{figure}

\begin{figure}[ht!]
\includegraphics[width=10cm]{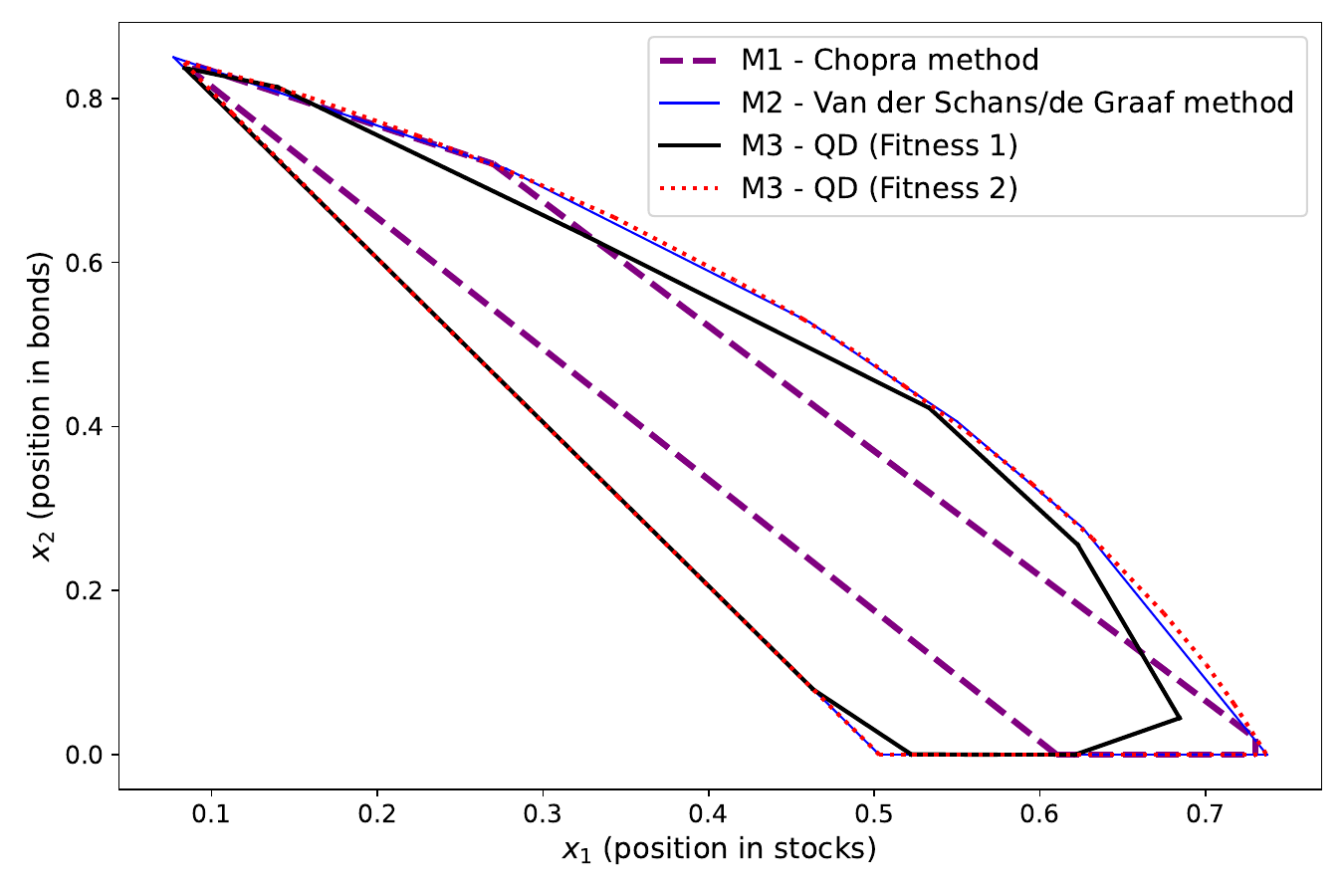}
\centering
\caption{Convex hulls in the genotypic/phenotypic space for different methods for a single QD run. The constraint $\sum x_i = 1$ introduces coplanarity in the otherwise three-dimensional space.}
\label{figure2}
\end{figure}

We begin by considering the toy example first introduced in \cite{chopra} and later revisited in \cite{degraaf}, in which only three ($N=3$) asset classes - stocks, bonds, and treasury bills are considered\footnote{More precisely, the assets include the S\&P 500 market index, Lehman Brothers Long Term Government Bond Index, and one-month Treasury bills. The original data is presented monthly and spans the period from 1980 to 1990, but the estimates are transformed into annual values in our work.}. The mean return, standard deviation, and correlation estimates for the three asset classes (given in Table \ref{table1}) are used to construct the expected return vector $\boldsymbol{\hat{\mu}}$ and the return covariance matrix $\boldsymbol{\hat{S}}$ estimates. The resulting MV efficient frontier is shown in Fig. \ref{figure1}, accompanied by additional elements, including the near-optimality region $\mathcal{R}$. To ensure consistency with \cite{degraaf}, the same reference portfolio $\boldsymbol{w_0}$ is adopted as well as the identical value of $c=0.1$. After running the QD algorithm $R=100$ times\footnote{At the start of each QD run, niches are recalculated, discarding the old CVT results.} for each fitness function, while using $\mathcal{B}_1$ and with $M=200$, the convex hulls depicting regions of near-optimality in the portfolio weights space are obtained. Each run takes less than $100$ seconds even without any parallelization. The maximum number of evaluations is set to $N_{max} = 250\ 000$ and the number of CVT samples equals $N_{CVT} = 10 \ 000$. Random initialization is done until $P_{init} = 10 \% $ of niches are filled with solutions. Fig. \ref{figure2} presents the results for four different methods (\textbf{M1} - Chopra \cite{chopra}, \textbf{M2} - van der Schans and de Graaf \cite{schans,degraaf}, \textbf{M3} - QD with \verb|Fitness1|, and \textbf{M4} - QD with \verb|Fitness2|), with more details given in Table \ref{table2}. Larger hull surfaces (volumes or hypervolumes in higher-dimensional spaces) are desirable, as they indicate greater compositional diversity within the set of found near-optimal portfolios, allowing investors more freedom to accommodate their specific preferences. The best results overall are obtained by \textbf{M4}, the modified version of \textbf{M3}, as it benefits from the additional diversification mechanism. The dominance of \textbf{M2} over \textbf{M3} with respect to the convex hull surface is not a matter of concern but rather an anticipated outcome. Namely, \textbf{M2} selects portfolios (exclusively from $\mathcal{R}$) by directly maximizing distances from the convex hull of previously found solutions, whereas \textbf{M3} seeks solutions within each niche whose risk profiles are strictly closest to $\mathcal{F}(\boldsymbol{w_0})$. As a result, \textbf{M3} pushes strongly towards $\mathcal{F}(\boldsymbol{w_0})$ (lying on the efficient frontier) regardless of whether exploring inside or outside of $\mathcal{R}$ and hence shrinks $\mathcal{S}$, the corresponding region in the risk-return space. We therefore suspect \textbf{M3} to yield portfolios with better risk-reward profiles (measured by the Sharpe ratio) than \textbf{M2} and also a lower $\mathcal{S}$ surface, all of which is indeed confirmed by the results laid out in Table \ref{table1}. It is also noteworthy that \textbf{M4} achieves an even higher value of the Sharpe ratio than \textbf{M3}, despite its emphasis on genotypic diversity. The likely reason is that, given that $\boldsymbol{\omega_0}$ is not the maximum Sharpe portfolio, it is possible for the neighboring risk profiles in $\mathcal{R}$ to dominate over it.

\begin{table}
\centering
\caption{The estimates from historical data (annual)}
\begin{tabular}{l|lllll|l}
\hline
& Mean (\%) & Std (\%) & Corr. & Corr. & Corr. & Optimal weights (\%) \\
&  & & stocks & bonds & T-Bills  & (moderate risk aversion) \\
\hline
Stocks & 15.876 & 16.603 & 1.000 & - & - & 58.1 \\
Bonds & 12.324 & 13.801 & 0.341 & 1.000 & - & 22.8 \\
T-Bills & 8.748 & 0.759 & -0.081 & 0.050  & 1.000 & 19.1 \\
\hline
\label{table1}
\end{tabular}
\end{table}

\begin{table}[ht!]
\centering
\caption{Performance evaluation of the four tested methods}
\begin{tabular}{|c|c|c|c|c|}
\hline
Method & Convex hull $\mathcal{H}$& Risk-return subspace $\mathcal{S}$ & Sharpe ratio & \% of niches \\
 & surface & surface ($\times 10^{3}$) & of portfolios & with opt. port. \\
\hline
\textbf{M1}\tablefootnote{Results from \cite{chopra} (deterministic).} & $0.0948$ & $0.1973$ & $1.1368 \pm 0.0959$ & - \\
\hline
\textbf{M2}\tablefootnote{Results from \cite{degraaf} (deterministic).} & $0.1724$ & $0.3712$ & $1.1762 \pm 0.1299$ & - \\
\hline
\textbf{M3} & $0.1529 \pm 0.0015 $ & $0.3652 \pm 0.0076$ & $1.2696 \pm 0.0009$ & $48.980 \pm 0.003 $ \\
\hline
\textbf{M4} & $\textbf{0.1746} \pm 0.0002$ & $0.3882 \pm 0.0008 $ & $\textbf{1.2824} \pm 0.0009$ & $\textbf{49.070} \pm 0.003$ \\
\hline
\end{tabular}
\label{table2}
\end{table}

\subsection{Higher-dimensional setting}

In this section, encouraged by the positive results from the toy example, we apply the proposed method to a higher-dimensional setting comprised of a large number of assets. Under such conditions, QD is expected to offer heuristic solutions in a reasonable time. More specifically, we consider the universe of $N=105$ different equity assets covering $L=11$ different sectors. 
The Sharpe optimal portfolio is used as the reference portfolio. Moreover, the Ledoit-Wolf shrinkage\cite{ledoit} is used\footnote{With constant variance set as the shrinkage target.} for estimating the covariance matrix $\boldsymbol{\hat{S}}$, while CAPM returns \cite{famafrench} are utilized to derive the mean return estimate $\boldsymbol{\hat{\mu}}$. The look-back window used for parameter estimation encompasses the period from January 2nd, 2020, to September 1st, 2023, corresponding to exactly $T=924$ trading days. We again set $c=0.01$, leading to a very strict criterion of near-optimality (i.e., a small surface of $\mathcal{R}$), and also select $M=5000$. 
Due to a plethora of assets, investors might prefer to first perform asset allocation at the sector/industry or some other macro level\footnote{Similar approaches are employed in top-down investment strategies such as Tactical Asset Allocation (TAA) \cite{faber}.} before selecting individual constituent securities. Consequently, we use the BS design $\mathcal{B}_2$. As the fitness function, $f_1$ is selected. The algorithm ultimately returns an archive of solutions $\mathcal{A}$ comprised of both near-optimal portfolios and non-near-optimal portfolios (for niches in which no near-optimal portfolios have been found).

Considering both the absence of suitable benchmarks and the computational challenges in calculating convex hull volumes in spaces with higher dimensionalities\footnote{With the QuickHull algorithm \cite{barber}, the execution time grows by $n^{\lfloor \frac{d}{2} \rfloor}$, where $n$ is the input size and $d$ the dimensionality.}, we draw inspiration from previous research on benchmarking QD algorithms \cite{flageat} and use it as a starting point in creating appropriate performance metrics. All of the used metrics are described in Table \ref{tab:metrics}. Remark that metrics $\mathcal{C'}$, $\operatorname{QDScore2}$ and $\operatorname{AP2}$ only consider niches in which near-optimal portfolios have been found, unlike $\operatorname{QDScore1}$ and $\operatorname{AP1}$, which (also) require information on other niches. Following the notation from Table \ref{tab:metrics}, the number of niches with and without near-optimal portfolios is $\mathcal{C'}M$ and $|\mathcal{A}|-\mathcal{C'}M$, respectively.

\begin{table}[htbp]
    \centering
    \caption{List of metrics}
    \label{tab:metrics}
    \begin{tabularx}{\textwidth}{>{\raggedright\arraybackslash}X >{\raggedright\arraybackslash}X >{\raggedright\arraybackslash}X}
        \toprule
        \textbf{Metric} & \textbf{Description} & \textbf{Expression} \\
        \midrule
        Modified coverage ($\mathcal{C'}$) & The proportion of niches with near-optimal portfolios  & $\frac{\text{No. of niches with n.o. portfolios}}{\text{No. of niches}}$ \\
        \midrule
        QD-score ($\operatorname{QDScore1}$) & The cumulative normalized fitness of all portfolios in $\mathcal{A}$ & $\sum_{i=1}^{|\mathcal{A}|} \frac{f_i-\min _j\left(f_j\right)}{\max _j\left(f_j\right)-\min _j\left(f_j\right)}$ \\
        \midrule
        Modified QD-score ($\operatorname{QDScoreMOD}$) & The cumulative normalized fitness of all found near-optimal portfolios  & $\sum_{i=1}^{\mathcal{C'}M} \frac{f_i-\min _j\left(f_j\right)}{\max _j\left(f_j\right)-\min _j\left(f_j\right)}$ \\
        \midrule
        Modified archive profile 1 ($\operatorname{AP1}$) & The proportion of found non-near-optimal portfolios exceeding some threshold value  & $\sum_{i=1}^{|\mathcal{A}|-\mathcal{C'}M} \mathbbm{1} \left(f_i \geq f_{\text {threshold }}\right)$ \\
        \midrule
        Modified archive profile 2 ($\operatorname{AP2}$) & The proportion of found near-optimal portfolios exceeding some threshold value  & $\sum_{i=1}^{\mathcal{C'}M} \mathbbm{1} \left(f_i \geq f_{\text {threshold }}\right)$ \\
        \bottomrule

    \end{tabularx}
\end{table}

\begin{figure}[ht!]
\includegraphics[width=12.5cm]{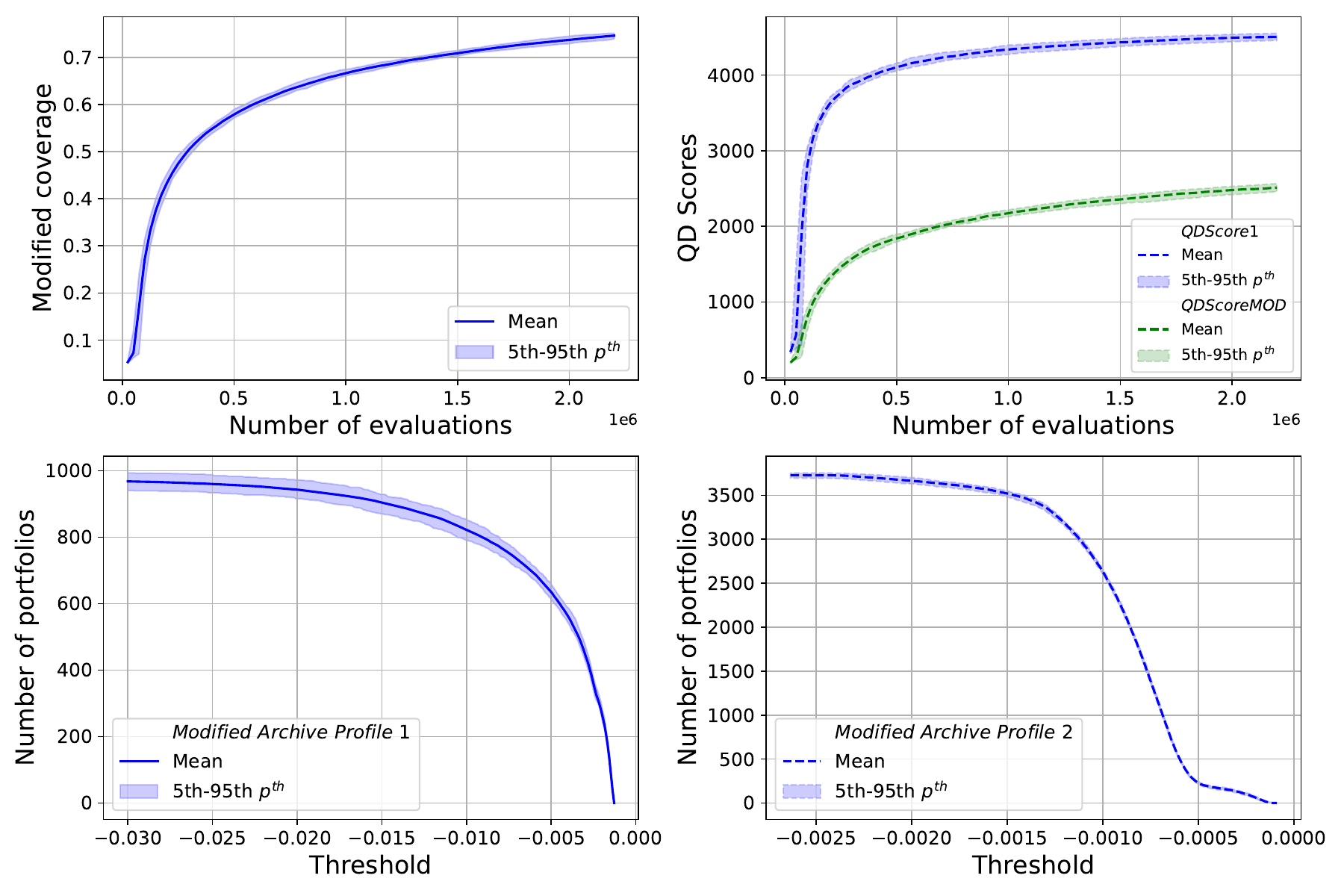}
\centering
\caption{The mean modified coverage $\mathcal{C'}$ (the upper left subplot), $\operatorname{QDScore1}$, and $\operatorname{QDScore2}$ (the upper right subplot) plotted against the number of evaluations, alongside empirical percentiles. The modified archive profiles are displayed at the bottom, with the set including non-near-optimal portfolios on the left and the set including near-optimal portfolios on the right.}
\label{figure4}
\end{figure}
\vspace{-0.50cm}
\subsubsection{Experimental results} Figure \ref{figure4} presents the main experimental results. The upper subplots show the first three metrics ($\mathcal{C'}$, QDScore1, QDScoreMOD) plotted against the number of QD evaluations, as well as the archive profiles (AP1 and AP2). The mean values as averaged over $R=20$ runs are shown, together with empirical percentiles (the $5$-th and the $95$-th $p^{th}$). As before, niches are recalculated from scratch at the beginning of each QD run. The maximum number of QD evaluations is set to $N_{max} = 2.2 \times 10^6$, $P_{init} = 10 \% $, and the number of CVT samples equals $N_{CVT} = 50 \ 000$. In terms of the modified coverage, it is evident that the percentage of niches with near-optimal solutions sharply rises, reaching $60 \%$ already at a bit over $500\ 000$ evaluations, and finally getting to approximately $75 \%$. Such high percentages clearly indicate that, despite the use of a stringent criterion for near-optimality, the method succeeds in finding a wide range of near-optimal portfolios encompassing various industry compositions and market capitalization values. This is a fortunate conclusion, especially for investors who prefer human-in-the-loop approaches, enabling them to incorporate their own preferences that might be hard to formalize, into the decision-making process. As for the QD-score metrics, we observe that $\operatorname{QDScore1}$ steadily increases over time, indicating an improvement in diversity/performance among the found near-optimal solutions (portfolios). Likewise, the $\operatorname{QDScore1}$ curve shows that the majority of niches are populated relatively quickly (in under $500\ 000$ evaluations). Also note that $\operatorname{QDScore1}$ expectedly converges faster than $\operatorname{QDScoreMOD}$, as filling niches with near-optimal solutions takes more time compared to arbitrary solutions. Modified archive profiles, calculated after performing the maximum number of QD evaluations, are provided at the bottom of Fig. \ref{figure4}. While AP1 and AP2 have similar shapes (under different threshold scales), a small right tail can be seen with AP2, showing a number of "super" near-optimal portfolios with risk profiles extremely similar to that of the reference portfolio. The significant compositional diversity of the obtained portfolios is depicted in Fig. \ref{figure3}, which displays the Sharpe optimal portfolio alongside two mutually diverse near-optimal portfolios generated by the method in the feature (behavior) space. Despite the multitude of potential solutions, in order to finalize the investment decision-making process, it is necessary to select a single portfolio from the set of obtained near-optimal portfolios. To this end, Alg. \ref{algo:invest} delineates the entire end-to-end investment decision-making process, incorporating the proposed QD method. 
\subsubsection{Robustness} 
\begin{figure}[ht!]
\includegraphics[width=11cm]{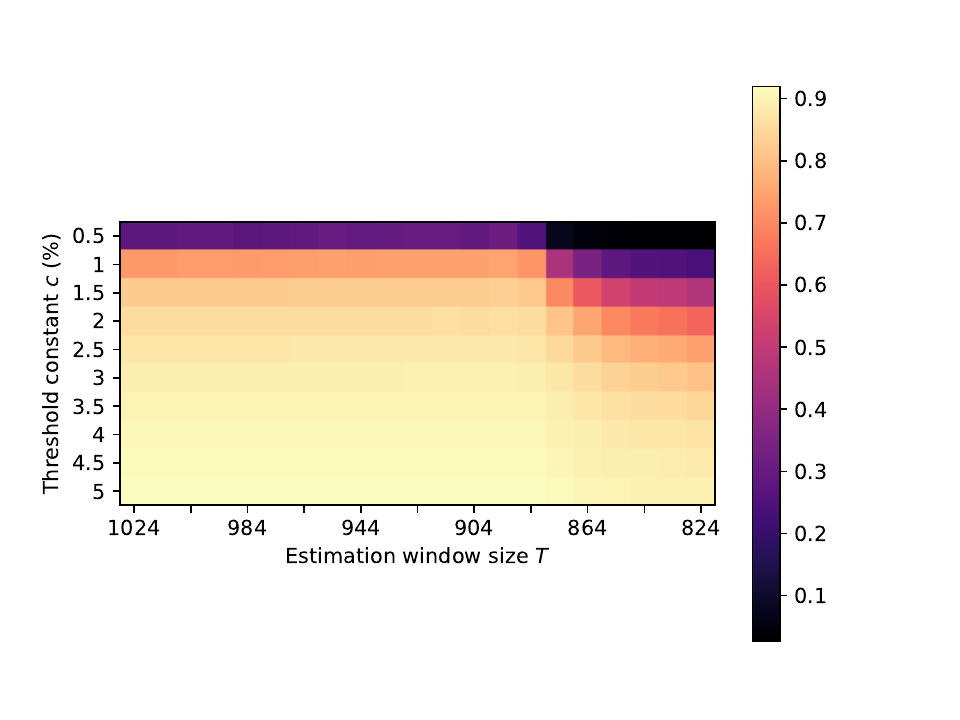}
\centering
\caption{The mean modified coverage $\mathcal{C'}$ for different threshold constants $c$ and estimation window sizes $T$. Observe the relatively high sensitivity of $\mathcal{C'}$ to the shortening of $T$, especially for stricter near-optimality criteria (i.e., for smaller $c$ values).}
\label{figure5}
\end{figure}

\begin{figure}[ht!]
\includegraphics[width=11cm]{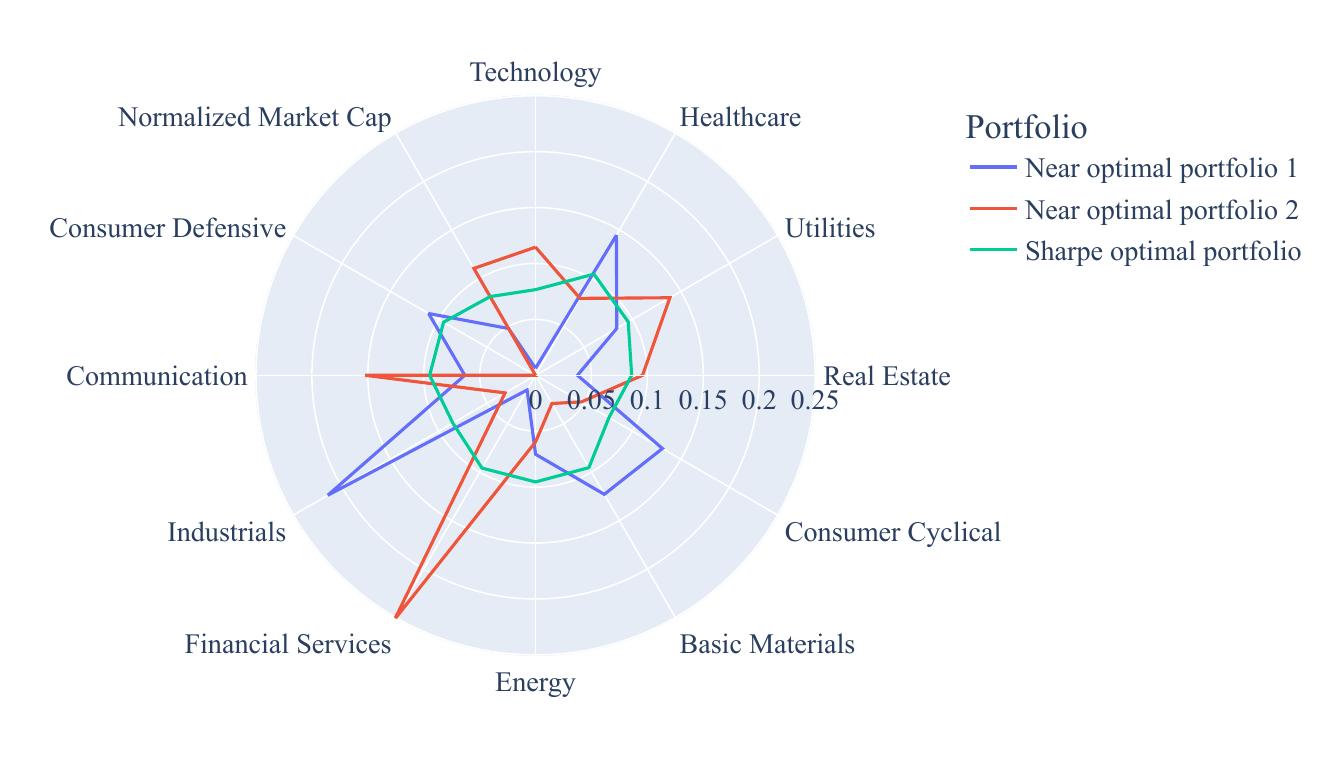}
\centering
\caption{Two of the obtained near-optimal portfolios juxtaposed against the Sharpe optimal portfolio in the BS. The first (second) near-optimal portfolio is highly concentrated in industrials (financial services) stocks and low (high) market cap stocks, while the Sharpe optimal portfolio remains more balanced.}
\label{figure3}
\end{figure}

Generally, the evaluation of QD solutions is exacerbated by the presence of stochasticity in the underlying environment \cite{digalakis}. In our case, the evaluation is fully deterministic once the estimates ($\boldsymbol{\hat{\mu}}$, $\boldsymbol{\hat{S}}$) are fixed. However, there is stochasticity involved due to the very fact that "true" parameter values ($\boldsymbol{\mu}$, $\boldsymbol{S}$) are hidden, whereas the estimates (which can be derived in multiple ways) represent random variables. With this in mind, we investigate the robustness of the generated portfolios to a certain type of change in the estimates ($\boldsymbol{\hat{\mu}}$, $\boldsymbol{\hat{S}}$). More specifically, we study whether previously found near-optimal portfolios generally remain near-optimal when reestimating under different estimation window\footnote{The choice of the estimation window size is a non-trivial issue that has been studied before \cite{gavsperov,wang2,wu}, with larger sizes leading to reduced estimation errors at the price of assuming unrealistically long stationarity periods.} sizes $T$. The results are shown in Fig. \ref{figure5}. As expected, the mean modified coverage $C'$ remains robust to changes in the estimation window size $T$ when larger threshold constants $c$ are employed. On the other hand, the sensitivity of $C'$ to changes in $T$, in particular to its reduction, is much more emphasized for $c$ values in the range $[0.5\%, 2.5\%]$, i.e., for stricter near-optimality criteria. For example, with $c=0.5 \%$ and $T = 824$, on average only $136.15$ niches (or $2.7\%$) contain near-optimal solutions. We leave for further research the study of whether solutions that remain near-optimal under a wider range of estimates present a superior investment choice.
\vspace{-0.8cm}

\begin{algorithm}[t!]
\SetAlgoLined
\kwInputs{$\text{Investor's risk aversion $\gamma$ and preferred BD $\boldsymbol{b}$, historical data $\mathcal{D}$} $}
\KwResult{$\text{Final portfolio} \  \boldsymbol{w}$}
$\boldsymbol{\hat{S}}, \boldsymbol{\hat{\mu}} \gets $ EstimateParameters($\mathcal{D}$) \\
$\boldsymbol{w_0} \gets$ CalculateEfficientFrontierPortfolio($\gamma$, $\boldsymbol{\hat{S}}$, $\boldsymbol{\hat{\mu}}$) \tcp*[r]{MV step - calculating the required portfolio on the efficient frontier}
$\mathcal{A} \gets$ RunQD \tcp*[r]{Run QD to obtain the archive of portfolios}
\text{$n_b \gets$ DetermineNicheIndex($\boldsymbol{b}$)} \tcp*[r]{Determine the niche index for the preferred BD}
\eIf {NearOptimalPortfolioExistsIn($n_b$)}{
    $\boldsymbol{w} \gets$ ElitePortfolioFrom($n_b$) 
}{
    \text{$n_b' \gets$ ClosestNicheWithNearOptimalPortfolio($n_b$)} \tcp*[r]{Among all niches with a near-optimal portfolio, select the one closest to $n_b$}
    $\boldsymbol{w} \gets$ ElitePortfolioFrom($n_b'$) 
}
\Return{$\boldsymbol{w}$}
\caption{Portfolio selection process}
\label{algo:invest}
\end{algorithm}

\vspace{-1.0cm}
\section{Conclusion and further work}

This paper is concerned with the problem of finding mutually diverse portfolios located in the region of near-optimality of some reference optimal portfolio. We introduce a novel method for discovering a wide spectrum of mutually diverse (in a predefined BS) near-optimal portfolios based on QD optimization. The main findings, pointing to high coverage and QD-scores, underscore the capacity of QD to serve as a novel instrument in the field of PO. In addition to QD, the \verb|knobelty| algorithm \cite{kelly} might be used to balance compositional diversity (\emph{novelty}) with proximity to the selected optimal portfolio. Similar approaches might also be employed for a somewhat different problem; to approximate the entire efficient frontier of optimal solutions, covering a wide range of different risk preferences and hence catering to a versatile set of investors. These include QD with a fitness function that penalizes distances from the efficient frontier in each niche, as well as vanilla NS formulations in which the risk-reward space is used as the phenotypic space. Other objectives besides MV may be explored in the future as well, together with different BS designs (e.g. factor-based, distance from the equally weighted or currently selected portfolio) and definitions of near-optimality (e.g. those employing soft constraints). Links with sparse PO should also be investigated. Lastly, we anticipate further work to leverage the power of QD in visualizing and illuminating the portfolio search space. More broadly, we hope to see future approaches harnessing the potential of open-endedness-based approaches, including QD and NS, in computational finance.



\bibliography{references}

\begin{thebibliography}{10}

\bibitem{sharpe}
William~F Sharpe.
\newblock The sharpe ratio.
\newblock {\em Streetwise--the Best of the Journal of Portfolio Management}, 3:169--85, 1998.

\bibitem{babcock}
Bruce~A Babcock, E~Kwan Choi, and Eli Feinerman.
\newblock Risk and probability premiums for cara utility functions.
\newblock {\em Journal of Agricultural and Resource Economics}, pages 17--24, 1993.

\bibitem{markowitz}
Harry~M Markowitz and G~Peter Todd.
\newblock {\em Mean-variance analysis in portfolio choice and capital markets}, volume~66.
\newblock John Wiley \& Sons, 2000.

\bibitem{best}
Michael~J Best and Robert~R Grauer.
\newblock On the sensitivity of mean-variance-efficient portfolios to changes in asset means: some analytical and computational results.
\newblock {\em The review of financial studies}, 4(2):315--342, 1991.

\bibitem{ledoit}
Olivier Ledoit and Michael Wolf.
\newblock Honey, i shrunk the sample covariance matrix.
\newblock {\em UPF economics and business working paper}, (691), 2003.

\bibitem{black}
Fischer Black and Robert Litterman.
\newblock Asset allocation: combining investor views with market equilibrium.
\newblock {\em Goldman Sachs Fixed Income Research}, 115(1):7--18, 1990.

\bibitem{demiguel}
Victor DeMiguel, Lorenzo Garlappi, Francisco~J Nogales, and Raman Uppal.
\newblock A generalized approach to portfolio optimization: Improving performance by constraining portfolio norms.
\newblock {\em Management science}, 55(5):798--812, 2009.

\bibitem{michaud}
Richard~O Michaud and Robert~O Michaud.
\newblock {\em Efficient asset management: a practical guide to stock portfolio optimization and asset allocation}.
\newblock Oxford University Press, 2008.

\bibitem{yin}
Chenyang Yin, Romain Perchet, and Fran{\c{c}}ois Soup{\'e}.
\newblock A practical guide to robust portfolio optimization.
\newblock {\em Quantitative Finance}, 21(6):911--928, 2021.

\bibitem{degraaf}
T~de~Graaf.
\newblock {\em Robust Mean-Variance Optimization}.
\newblock PhD thesis, Master Thesis, Leiden University \& Ortec Finance, 2016.

\bibitem{schans}
Martin van~der Schans and Tanita de~Graaf.
\newblock Robust optimization by constructing near-optimal portfolios.
\newblock {\em Available at SSRN 3057258}, 2017.

\bibitem{wang}
Lipo Wang.
\newblock {\em Support vector machines: theory and applications}, volume 177.
\newblock Springer Science \& Business Media, 2005.

\bibitem{wales}
David~J Wales and Jonathan~PK Doye.
\newblock Global optimization by basin-hopping and the lowest energy structures of lennard-jones clusters containing up to 110 atoms.
\newblock {\em The Journal of Physical Chemistry A}, 101(28):5111--5116, 1997.

\bibitem{eeghen}
WJB van Eeghen, OW~van Gaans, and M~van~der Schans.
\newblock Analysis of near-optimal portfolio regions and polytope theory.
\newblock 2018.

\bibitem{cajas}
Dany Cajas.
\newblock Robust portfolio selection with near optimal centering.
\newblock {\em Available at SSRN 3572435}, 2019.

\bibitem{chopra}
Vijay~Kumar Chopra.
\newblock Improving optimization.
\newblock {\em The Journal of Investing}, 2(3):51--59, 1993.

\bibitem{benita}
Golan Benita, Nadine Baudot-Trajtenberg, and Amit Friedman.
\newblock The challenges of managing large fx reserves: the case of israel.
\newblock {\em BIS Paper}, (104m), 2019.

\bibitem{fagerstrom}
Sixten Fagerstr{\"o}m and Gustav Oddshammar.
\newblock Portfolio optimization-the mean-variance and cvar approach.
\newblock 2010.

\bibitem{brabazon}
Anthony Brabazon, Michael O'Neill, and Ian Dempsey.
\newblock An introduction to evolutionary computation in finance.
\newblock {\em IEEE Computational Intelligence Magazine}, 3(4):42--55, 2008.

\bibitem{branke}
J{\"u}rgen Branke, Benedikt Scheckenbach, Michael Stein, Kalyanmoy Deb, and Hartmut Schmeck.
\newblock Portfolio optimization with an envelope-based multi-objective evolutionary algorithm.
\newblock {\em European Journal of Operational Research}, 199(3):684--693, 2009.

\bibitem{qi}
Rongbin Qi and Gary~G Yen.
\newblock Hybrid bi-objective portfolio optimization with pre-selection strategy.
\newblock {\em Information Sciences}, 417:401--419, 2017.

\bibitem{chatz}
Konstantinos Chatzilygeroudis, Antoine Cully, Vassilis Vassiliades, and Jean-Baptiste Mouret.
\newblock Quality-diversity optimization: a novel branch of stochastic optimization.
\newblock In {\em Black Box Optimization, Machine Learning, and No-Free Lunch Theorems}, pages 109--135. Springer, 2021.

\bibitem{lehman}
Joel Lehman and Kenneth~O Stanley.
\newblock Novelty search and the problem with objectives.
\newblock {\em Genetic programming theory and practice IX}, pages 37--56, 2011.

\bibitem{gomes2013evolution}
Jorge Gomes, Paulo Urbano, and Anders~Lyhne Christensen.
\newblock Evolution of swarm robotics systems with novelty search.
\newblock {\em Swarm Intelligence}, 7:115--144, 2013.

\bibitem{pugh2016quality}
Justin~K Pugh, Lisa~B Soros, and Kenneth~O Stanley.
\newblock Quality diversity: A new frontier for evolutionary computation.
\newblock {\em Frontiers in Robotics and AI}, 3:40, 2016.

\bibitem{zhang}
Tianping Zhang, Yuanqi Li, Yifei Jin, and Jian Li.
\newblock Autoalpha: An efficient hierarchical evolutionary algorithm for mining alpha factors in quantitative investment.
\newblock {\em arXiv preprint arXiv:2002.08245}, 2020.

\bibitem{yuksel}
Kamer~Ali Yuksel.
\newblock Generative meta-learning robust quality-diversity portfolio.
\newblock In {\em Proceedings of the Companion Conference on Genetic and Evolutionary Computation}, pages 787--790, 2023.

\bibitem{vassila}
Vassiiis Vassiliades and Jean-Baptiste Mouret.
\newblock Discovering the elite hypervolume by leveraging interspecies correlation.
\newblock In {\em Proceedings of the Genetic and Evolutionary Computation Conference}, pages 149--156, 2018.

\bibitem{digalakis}
Jason~G Digalakis and Konstantinos~G Margaritis.
\newblock On benchmarking functions for genetic algorithms.
\newblock {\em International journal of computer mathematics}, 77(4):481--506, 2001.

\bibitem{bossens}
David~M Bossens and Danesh Tarapore.
\newblock Quality-diversity meta-evolution: customising behaviour spaces to a meta-objective.
\newblock {\em arXiv preprint arXiv:2109.03918}, 2021.

\bibitem{sfikas}
Konstantinos Sfikas, Antonios Liapis, and Georgios~N Yannakakis.
\newblock Monte carlo elites: Quality-diversity selection as a multi-armed bandit problem.
\newblock In {\em Proceedings of the Genetic and Evolutionary Computation Conference}, pages 180--188, 2021.

\bibitem{vassiliades}
Vassilis Vassiliades, Konstantinos Chatzilygeroudis, and Jean-Baptiste Mouret.
\newblock Using centroidal voronoi tessellations to scale up the multidimensional archive of phenotypic elites algorithm.
\newblock {\em IEEE Transactions on Evolutionary Computation}, 22(4):623--630, 2017.

\bibitem{mouret}
Jean-Baptiste Mouret and Jeff Clune.
\newblock Illuminating search spaces by mapping elites.
\newblock {\em arXiv preprint arXiv:1504.04909}, 2015.

\bibitem{famafrench}
Eugene~F Fama and Kenneth~R French.
\newblock The capital asset pricing model: Theory and evidence.
\newblock {\em Journal of economic perspectives}, 18(3):25--46, 2004.

\bibitem{faber}
Meb Faber.
\newblock A quantitative approach to tactical asset allocation.
\newblock {\em The Journal of Wealth Management, Spring}, 2007.

\bibitem{barber}
C~Bradford Barber, David~P Dobkin, and Hannu Huhdanpaa.
\newblock Qhull: Quickhull algorithm for computing the convex hull.
\newblock {\em Astrophysics Source Code Library}, pages ascl--1304, 2013.

\bibitem{flageat}
Manon Flageat, Bryan Lim, Luca Grillotti, Maxime Allard, Sim{\'o}n~C Smith, and Antoine Cully.
\newblock Benchmarking quality-diversity algorithms on neuroevolution for reinforcement learning.
\newblock {\em arXiv preprint arXiv:2211.02193}, 2022.

\bibitem{gavsperov}
Bruno Ga{\v{s}}perov, Fredi {\v{S}}ari{\'c}, Stjepan Begu{\v{s}}i{\'c}, and Zvonko Kostanj{\v{c}}ar.
\newblock Adaptive rolling window selection for minimum variance portfolio estimation based on reinforcement learning.
\newblock In {\em 2020 43rd International Convention on Information, Communication and Electronic Technology (MIPRO)}, pages 1098--1102. IEEE, 2020.

\bibitem{wang2}
Pei-Ting Wang and Chung-Han Hsieh.
\newblock On data-driven log-optimal portfolio: A sliding window approach.
\newblock {\em IFAC-PapersOnLine}, 55(30):474--479, 2022.

\bibitem{wu}
Chuanzhen Wu.
\newblock Window effect with markov-switching garch model in cryptocurrency market.
\newblock {\em Chaos, Solitons \& Fractals}, 146:110902, 2021.

\bibitem{kelly}
Jonathan Kelly, Erik Hemberg, and Una-May O’Reilly.
\newblock Improving genetic programming with novel exploration-exploitation control.
\newblock In {\em Genetic Programming: 22nd European Conference, EuroGP 2019, Held as Part of EvoStar 2019, Leipzig, Germany, April 24--26, 2019, Proceedings 22}, pages 64--80. Springer, 2019.

\end{thebibliography}

%
%
%
%

\end{document}